\documentclass[twocolumn,aps,floatfix]{revtex4}
\usepackage{graphicx,epsfig,subfig,dcolumn,bm,mathrsfs,amsmath,color,setspace}
\begin{document}

\title{AC-field-induced Polarization for Uncharged Colloids in Salt Solution: \\A Dissipative Particle Dynamics Simulation}

\author{Jiajia Zhou}
\email[]{zhou@uni-mainz.de}
\altaffiliation{}
\affiliation{Institut f\"ur Physik, Johannes Gutenberg-Universit\"at Mainz\\
 Staudingerweg 7, D-55099 Mainz, Germany} 
\author{Friederike Schmid}
\email[]{friederike.schmid@uni-mainz.de}
\affiliation{Institut f\"ur Physik, Johannes Gutenberg-Universit\"at Mainz\\
 Staudingerweg 7, D-55099 Mainz, Germany} 


\begin{abstract}
We study the response of a spherical colloid under alternating electric fields (AC-fields) by mesoscopic simulation method, accounting in full for hydrodynamic and electrostatic interactions.
We focus on a special case of uncharged colloids.
The main polarization mechanism is the ``volume-polarization'', where ionic currents are deflected by the core of the uncharged colloid. 
Specifically, we compute the polarizability of a single colloid and systematically investigate the effect of AC field frequency and salt concentration.
The simulation results are compared with predictions from classical Maxwell-Wagner theory and electrokinetic theory. 
\end{abstract}


\maketitle

\section{Introduction}
\label{sec:introduction}

Colloids suspended in an aqueous solution respond to external fields on relatively short time scales and in an often fully reversible way.
This makes the application of electric fields one of the most attractive methods to manipulate single colloid or colloidal suspensions. 
Electrophoresis, the translation of individual colloid under a static external field \cite{RSS}, is widely used to measure the surface charge density of colloidal particles. 
An alternating electric field (AC-field) can also be used to probe the time-dependent properties of the colloidal suspension and provide substantially more information about the colloids than the static field.

Let us consider a simple situation where a spherical colloid is immersed in either a dielectric or conducting medium such as a salt solution. 
The particle is often negatively charged, either by ionization or dissociation of a surface group, or by preferential adsorption of ions from the solution.
The surface charges result in the formation of an electrical double layer (EDL) around the particle. 
The inner layer is called Stern layer and consists of ions which are absorbed on the particle surface. 
The outer layer is called diffuse layer and consists of counterions accumulated around the particle surface due to the electrostatic attraction between opposite charges. 
The ions in the diffuse layer are mobile and experience thermal motion. 
In Debye-H\"uckel approximation, the thickness of the diffuse layer is characterized by the Debye length
\begin{equation}
  \label{eq:Debye}
  l_D = \kappa^{-1} = \left[ \frac{ \sum_i z_i^2 e^2 \rho_i(\infty) }{ \epsilon_m k_B T} \right]^{-\frac{1}{2}},
\end{equation}
where the sum runs over different ion types, $z_i$ and $\rho_i(\infty)$ are the valence and bulk concentration for $i$-th ion, respectively, and $\epsilon_m$ is the medium permittivity.   

When an electric field of the form ${\bf E}={\bf E}_0 \exp(i\omega t)$ is applied to the suspension, both the colloidal particle and the electric double layer are polarized.
The colloid acquires a dipole moment of the form $\mathbf{p} \exp(i \omega t)$, and the amplitude of the dipole moment can be written as
\begin{equation}
  {\bf p} = \alpha(\omega) {\bf E}_0 ,
\end{equation}
where $\alpha(\omega)$ is the polarizability of the particle. 
The polarizability is used to characterize the colloid response to the external fields, and it is a complex function of the frequency and the amplitude of the external field. 
For weak fields, the dipole moment is proportional to the magnitude of the external field; thus the polarizability does not depend on the field strength.  
Various effects contribute to the dipole moment, which we will discuss in detail in the following.

When no external fields are present, the diffuse layer of a charged colloid is spherically symmetric. 
Without loss of generality, we assume that the colloid is negatively charged (and hence the counterions are positively charged cations).
The center of the counterion cloud and the colloid center coincide, resulting in a zero dipole moment.
When an external electric field is applied, the negatively charged colloid moves in the direction opposite to the field, while the positive counterions move in the direction of the field.
The counterions move in and out of the electrical double layer, but when averaged over time, the counterion cloud has a dynamically stable ellipsoidal shape.
The center of mass of the counterions is displaced with respect to the colloid center in the direction of the field, which results in a net dipole moment that points in the same direction.
This type of polarization is referred to as \emph{field-polarization}. 
This mechanism can also be viewed as ion fluxes being deflected by the electric field produced by the charged colloid, as in ref. \cite{Dhont2010}.

\begin{figure}[htbp]
  \centering
  \includegraphics[width=1.0\columnwidth]{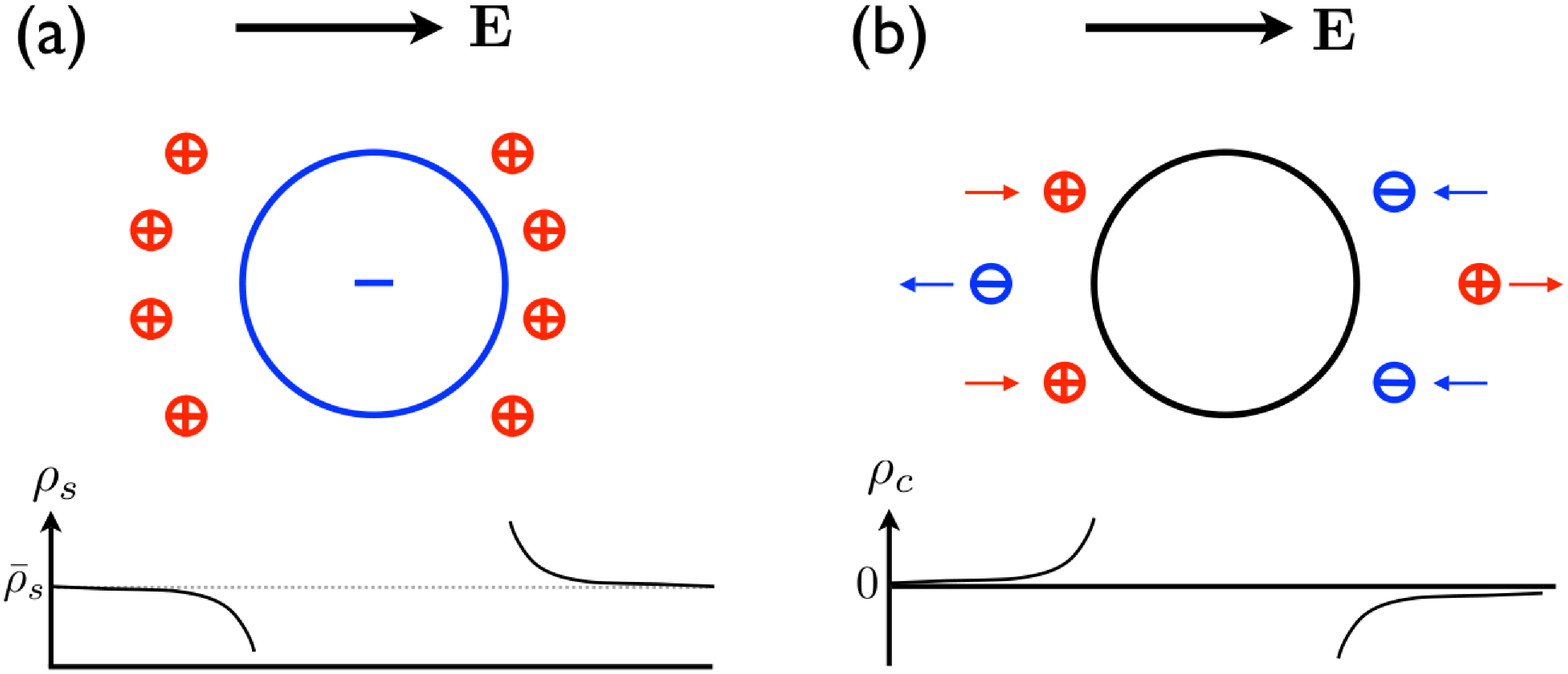}
  \caption{(Colour on-line) The polarization mechanism due to the obstacle effect of the colloid core. Different interpretations exist for charged and uncharged colloids. (a) ``concentration polarization'' for a negatively charged colloid. The positive counterion cloud shifts to the left and the effective dipole moment points in the direction opposite to the external field. The bottom shows the electrolyte concentration profile $\rho_s$ along a line passing through the colloid center. (b) ``volume-polarization'' for an uncharged colloid. The bottom shows the charge density profile $\rho_c=e(\rho_+-\rho_-)$, where $\rho_+$ and $\rho_-$ are the densities for cations and anions, respectively.}
  \label{fig:intro}
\end{figure}

For field-polarization to take effect, the charged components should be mobile. 
The dipole moment is produced due to the fact that oppositely charged components move in different directions in an external field. 
There exists another mechanism of polarization which can be traced down to the presence of the colloidal particle as an obstacle that microions cannot penetrate. 
This mechanism has been described in the literature under different names, and for charged colloids, it is often referred to as \emph{concentration polarization} \cite{Shilov2000,DukhinShilov}.
Fig. \ref{fig:intro}(a) shows a schematic description of the mechanism. 
On the right-hand side of the colloid, a flux of negatively charged ions moves in the direction opposite to the field and hits the impenetrable core of the colloid.  
Combined with the abundant positive ions near the surface, the accumulation of negative ions leads to an increase in the neutral electrolyte concentration.
The slightly enhanced electrolytes produce a locally reduced Debye length, such that the counterions are compressed close to the surface. 
A similar process on the left-hand side leads to a decrease in electrolyte concentration and expansion of the counterion cloud. 
The net effect is a shift to the left of the center of the counterion cloud, and the resulting dipole moment points anti-parallel to the field. 

The polarization mechanism due to the colloid obstacle is not restricted to charged particles.  
For uncharged particles, a dipole moment can also be induced by the movement of the salt ions.
This is demonstrated in Fig. \ref{fig:intro}(b).  
The positive ions move in the direction of the electric field, and tend to accumulate on the left-hand side of the colloid due to the impenetrable core of the colloid. 
This produces a slight increase of the cation concentration on the left-hand side of the colloid. 
A similar effect occurs on the right-hand side of the colloid where negative ions accumulate, resulting in an increase of anion concentration. 
The combined effect again leads to an effective dipole moment which points in the direction opposite to the external field. 
This is referred to as \emph{volume-polarization} \cite{Dhont2010}. 
In this case, a variation of the charge density is produced near the colloid surface, whereas in the charged case, it is a variation of electrolyte density.  

The relative importance of these two types of polarization mechanism depends on the bare surface charges of the colloid, the bulk salt concentration, and the frequency and amplitude of the external field. 
It is easy to see that the field-polarization is only present when the colloids are charged. 
For charged particle, the situation is more complicated than for uncharged ones due to the presence of the electric double layer. 
Other effects also contribute to the polarization for charged particles, such as the lateral movement of microions in the Stern layer \cite{Mangelsdorf1998a,Mangelsdorf1998b,Carrique2007}, and flow-induced effect due to the electrophoretic motion of the colloid \cite{Mangelsdorf1992,ZhaoHui2009}.
The situation is comparably simple for uncharged colloids, as the volume-polarization is the only mechanism for polarization.

In this work, we study the case of uncharged colloids and focus on the effect of volume-polarization. 
This work is motivated by recent theoretical results due to Dhont and Kang \cite{Dhont2010}, who gave accurate predictions for various response functions of uncharged colloids in salt solutions.
We use Dissipative Particle Dynamics (DPD) simulations \cite{Koelman1993,Espanol1995,Groot1997} to study the dielectric response of a spherical colloid under AC fields, including in full the hydrodynamic and electrostatic interactions. 
In a recent publication \cite{2012_ac}, we have introduced a DPD model for colloids with no-slip surface, and presented preliminary results on the frequency-dependent dielectric response of nanoscale colloids.
The results were in surprisingly good agreement with the prediction of the Maxwell-Wagner theory, even though the latter was originally devised for micro-sized colloids. 
In the present work, we investigate the reason for this surprising agreement, and analyze in details the different physical factors that contribute to the polarizability of the colloid. 
To this end, we systematically vary the salt concentration, and we examine the contribution of the microions in different ion shells to the dielectric response. 
We compare our simulation results with a recent electrokinetic theory by Dhont and Kang \cite{Dhont2010} and find very good agreement. 
The remainder of this article is organized as follows: in section \ref{sec:theory} we summarize the main theoretical results for uncharged colloids, which we will compare with our simulation results. 
In section \ref{sec:model}, we give a brief introduction of the simulation model and describe important parameters of the system. 
We present the simulation results on the polarizability and effects of external AC fields in section \ref{sec:results}. 
Finally, section \ref{sec:summary} concludes with a brief summary.

\section{Theories for uncharged colloids}
\label{sec:theory}

In this section, we briefly review two types of theories for uncharged colloids: the Maxwell-Wagner theory and the theory by Dhont and Kang \cite{Dhont2010}, which is based on the standard electrokinetic approach. 
We present only the important results in the main text and refer to the Appendices for detailed derivations. 

We consider a spherical particle of radius $R$, made of a material with permittivity $\epsilon_p$ and conductivity $K_p$, immersed in a medium with permittivity $\epsilon_m$ and conductivity $K_m$.
In the simulation, we will simplify the picture by considering only non-conducting colloids ($K_p=0$) and assuming that the particle and its surrounding medium have the same permittivity ($\epsilon_p=\epsilon_m=\epsilon$). 

The classical Maxwell-Wagner theory \cite{Maxwell1954,Wagner1914} can be applied when there is a material property mismatch for the two components forming an interface.
The condition of continuity of the normal components of the displacement vector and the current density vector at the interface implies
\begin{equation}
  \label{eq:ekek}
  \frac {\epsilon_m}{K_m} = \frac {\epsilon_p}{K_p}.
\end{equation}
If condition (\ref{eq:ekek}) is not satisfied, free ionic charges accumulate at the interface. 
The polarizability for a non-conducting colloid in a permittivity-matched medium (see Appendix \ref{app:mw}) can be written as 
\begin{equation}
  \label{eq:alpha_mw}
  \alpha(\omega) = -2\pi \epsilon R^3 \frac{ 1 - i (\omega / \omega_{\rm mw}) } { 1 + (\omega / \omega_{\rm mw})^2 }
\end{equation}
with 
\begin{equation}
  \label{eq:w_mw}
  \omega_{\rm mw}= \frac{2K_m}{3\epsilon}. 
\end{equation}
The Maxwell-Wagner time $\tau_{\rm mw}=1/\omega_{\rm mw}$ corresponds to the time scale for the free charge formation at the interface.  

In the Maxwell-Wagner theory, the particle and its surrounding medium are treated as homogeneous substances, and the induced charges only appear on the particle-fluid interface.
The theory takes into account only the bulk properties of the components and neglects the effect of spacial distribution of the polarization charges. 
This simplification is valid for large particles, where the length scale for the variation is much less than the colloid size. 
For nanometer-size particles, the distribution of the microions near the interface may become important.  

The theory taking into consideration of the effect of space charge variation is based on the well-known, standard electrokinetic equations \cite{RSS}.
Specifically, the theory for uncharged colloids describes the interplay of ion concentration profiles and the electric potential. 
Two differential equations are considered: the first one is the Poisson equation, which connects the charge density to the Laplacian of the potential; the second one is the Nernst-Planck equation, which relates the time derivative of the ion concentration to the potential and flow velocity. 
A third equation, the Navier-Stokes equation, connects the flow velocity with the electrostatic force density.
For the uncharged case, the convective contribution from the flow is a second-order effect and the Navier-Stokes equation can be omitted. 
The two equations (Poisson and Nernst-Planck) can then be solved with suitable boundary condition at the particle-fluid interface. 

In the literature, the majority of the studies have dealt with the general case of charged particles. 
For uncharged particles, Bonincontro {\it et al.} \cite{Bonincontro1980} obtained solutions for the electrokinetic equations and studied the effect of volume-polarization for a sphere in an 1-1 electrolyte solution. 
However, their result for the dipole moment differs from the Maxwell-Wagner theory in the limit of $\kappa R \rightarrow \infty$. 
Later Garcia {\it et al.} \cite{Garcia1985} derived a formula with the correct asymptotic behavior and showed that the effect of space distribution of charges leads to a transition frequency higher than the Maxwell-Wagner results.
Recently, Dhont and Kang have calculated the polarizability for an uncharged spherical colloid \cite{Dhont2010}:
\begin{equation}
  \label{eq:alpha_dhont}
  \alpha(\omega) = -4\pi\epsilon R^3 \frac{\kappa^2}{s^2} \: 
  \frac { 1 + sR + \frac{1}{3} (sR)^2 } { 2 + 2 sR + (sR)^2 - \frac{1}{3} (\kappa R)^2 }, 
\end{equation}
(see Appendix \ref{app:dhont} for notations). 
They also derived explicit expressions of the response functions for the polarization charge density and potential.
Their results indicate that the characteristic length for the polarization charge variation is of the order of $l_D$.

\section{Simulation Model}
\label{sec:model}

In this section, we briefly review our simulation model for a colloidal particle in a salt solution and describe some important physical quantities. 
A more detailed description can be found in ref. \cite{2012_ac}.

Our simulation system has three components: the solvents, the microions, and the colloidal particle. 
The solvent is modeled as a fluid of DPD beads, where DPD is used as a canonical thermostat ({\it i.e.} including the dissipative and stochastic part) without conservative forces \cite{Soddemann2003}. 
In the following, physical quantities will be reported in a model unit system of $\sigma$ (length), $m$ (mass), $\varepsilon$ (energy), $e$(charge) and derived time unit $\tau=\sigma\sqrt{m/\varepsilon}$. 
In these units, the temperature of the system is $k_BT=1.0\,\varepsilon$, the density of the fluid is $3.0\, \sigma^{-3}$, and each solvent bead has a mass $m$. 
The DPD friction coefficient is set to $\gamma_{DPD}=5.0 \, \sqrt{m\varepsilon}/\sigma$ and the cutoff radius is $r_c = 1.0\,\sigma$.
The shear viscosity is measured using the method described in ref. \cite{Smiatek2008} by analyzing a plane Poiseuille flow in a microchannel. 
For our system, we obtained a shear viscosity $\eta_s=1.23 \pm 0.01\, m/\tau$, which is in good agreement with ref. \cite{Smiatek2009}.  

Salt microions are introduced in the system as pairs of positively and negatively charged beads. 
We only consider the monovalent case where salt ions carry a single elementary charge  $\pm e$. 
The interactions between microions have two parts: one is the electrostatic interactions with the Bjerrum length $l_B = e^2/(4\pi\epsilon_m k_BT)$ of the fluid set to $1.0\,\sigma$.
The other interaction is a short-range repulsive Weeks-Chandler-Andersen interaction \cite{WCA},
\begin{equation}
  \label{eq:WCA}
  V_{WCA}(r) = \left\{ \begin{array}{ll}
      4\varepsilon \left[ ( \frac{\sigma}{r-r_0} )^{12} 
           - ( \frac{\sigma}{r-r_0} )^6 + \frac{1}{4} \right] \quad & \mbox{for } \quad r<r_c \\
       0 & \mbox{otherwise}
      \end{array} \right.
\end{equation}
which prevents the collapse of charged system.
The cutoff radius is set at the potential minimum $r_c=r_0 + \sqrt[6]{2} \,\sigma$.
The microions have a size of $1.0\,\sigma$ ($r_0=0$). 

\begin{figure}[htbp]
  \centering
  \includegraphics[width=0.6\columnwidth]{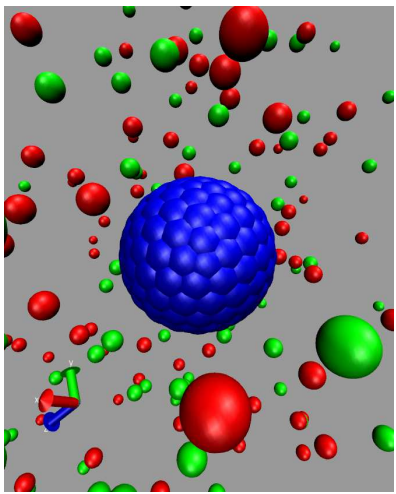}
  \caption{(Colour on-line) Snapshot of an uncharged colloidal particle in a salt solution. The surface sites are represented by the blue beads. The red and green beads are cations and anions, respectively. Solvent beads are not shown here.}
  \label{fig:snap}
\end{figure}

The colloidal particle is represented by a large sphere with interacting sites on its surface \cite{Lobaskin2004,Lobaskin2004a,Lobaskin2007,Chatterji2005,Chatterji2007,Giupponi2011}. 
The short-range repulsive interaction between the colloid and fluid/ microion particles is modeled through a similar WCA  potential (\ref{eq:WCA}).
The radius of the colloid is $R=r_0+\sigma=3.0\, \sigma$. 
To implement no-slip boundary conditions at the surface, a set of $N$ interaction sites is distributed evenly on the surface.
The interaction between these surface sites and the solvent beads is modeled using DPD, with the same DPD friction constant as that between fluid beads, but twice the cutoff range to ensure that a sufficient number of solvent particle can interact with the colloid.
The mass of the colloidal particle is $M=100\, m$, and the moment of inertia is $I=360\, m\sigma^2$, corresponding to a sphere with constant volume density. 
Fig. \ref{fig:snap} shows a representative snapshot of a single colloidal particle in a salt solution. 

We carried out simulations using the open source package ESPResSo \cite{ESPResSo}. 
Modifications have been made to incorporate an external time-dependent electric field. 
A cubic simulation box of $L=30\,\sigma$ with periodic boundary conditions in all three directions was used for all simulations. 
Electrostatic interactions are calculated using Particle-Particle-Particle Mesh (P3M) method \cite{HockneyEastwood,Deserno1998,Deserno1998a}.  
A time step of $\Delta t = 0.01\, \tau$ is used for the integration.

\begin{figure}[htbp]
  \centering
  \includegraphics[width=0.9\columnwidth]{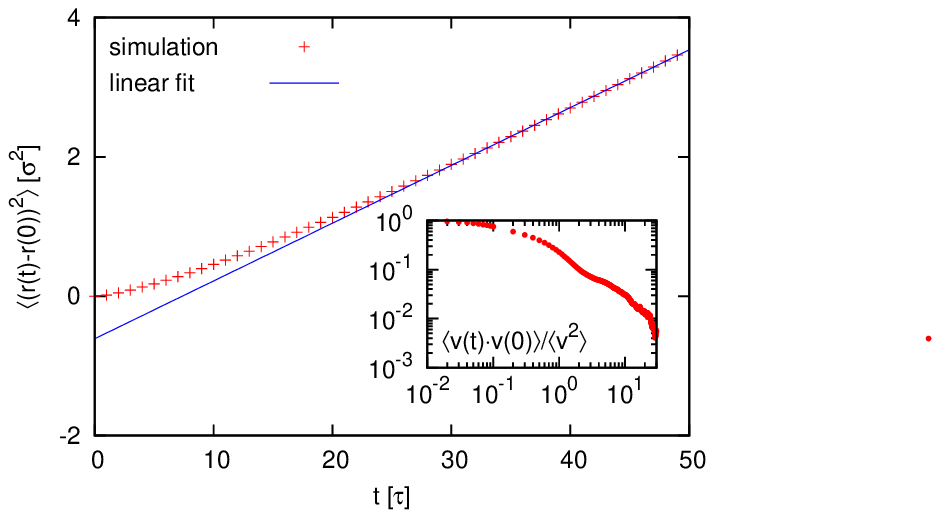}
  \caption{The mean-square displacement of an uncharged colloidal particle with radius $R=3.0\,\sigma$ in a salt-free solution. The linear regression is done for $t>30\, \tau$. Inset: normalized velocity autocorrelation function for the colloid, from ref. \cite{2012_ac}.}
  \label{fig:r2}
\end{figure}

As a benchmark test of our colloid model, we have performed simulations of an uncharged colloid in a salt-free solution. 
The electric field is turned off.
We measured the mean-square displacement $\langle ( \mathbf{r}(t)-\mathbf{r}(0) )^2 \rangle$ of the colloid as a function of time, shown in Fig. \ref{fig:r2}. 
Ballistic motion is observed for small times, as the mean-square displacement scales $\sim t^2$. 
At late time, the colloidal particle experienced many random collisions with the solvent beads, leading to a linear dependency on time. 

The diffusion constant of the colloid can be determined by a linear regression at late times,
\begin{equation}
  D = \lim_{t\rightarrow \infty} \frac{ \langle (\mathbf{r}(t)-\mathbf{r}(0))^2 \rangle }{ 6t }.
\end{equation}
The result after averaging five independent runs is $D=0.014 \pm 0.002 \, \sigma^2/\tau$.  
Alternatively, the diffusion constant can also be determined from the velocity autocorrelation function, using the Green-Kubo relation
\begin{equation}
  D = \frac{1}{3} \int_0^{\infty} {\rm d}t \langle \mathbf{v}(t) \cdot \mathbf{v}(0) \rangle.
  \label{eq:GK}
\end{equation}
The velocity autocorrelation function has already been determined in our previous simulations \cite{2012_ac} and is shown in the inset of Fig. \ref{fig:r2}.
The integration Eq. (\ref{eq:GK}) gives the diffusion constant $D=0.013 \pm 0.002\, \sigma^2/\tau$, in good agreement with the result from the mean-square displacement. 
For comparison, the diffusion constant of a Stokes sphere of radius $R=3.0 \, \sigma$ is $D=k_BT/(6\pi \eta_s R) = 0.0144 \pm 0.0002\, \sigma^2/\tau$, which indicates that the no-slip boundary condition is achieved in our colloid model.

\begin{figure}[htbp]
  \centering
  \includegraphics[width=0.9\columnwidth]{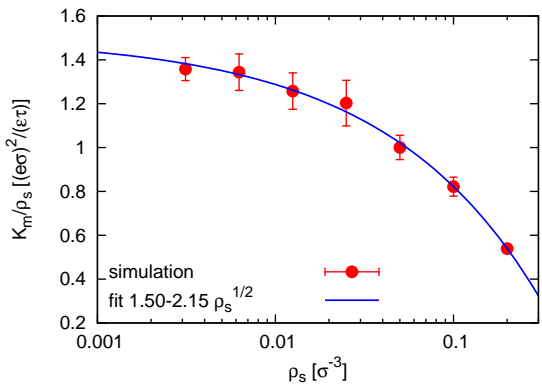}
  \caption{The molar conductivity $K_m/\rho_s$, or the scaled diffusion constant of microions $(2 e^2/k_BT) D_I$, as a function of salt concentration $\rho_s$. The curve is a fit to Kohlrausch law with fitting parameters $A=1.50$ and $B=2.15$.}
  \label{fig:Km}
\end{figure}

One important quantity required for comparison with the theory is the conductivity of the solution $K_m$, which is related to the diffusion constant $D_I$ for microions
\begin{equation}
  \label{eq:Km}
  K_m = \frac{ 2 e^2 \rho_s D_I }{ k_BT }.
\end{equation}
The formula is applied to 1-1 electrolyte solution and assumes that cations and anions have the same diffusion constant. 
The diffusion constant $D_I$ can be determined by measuring the mean-square displacement for microions. 
We performed simulations with different salt concentrations, varying from $\rho_s=0.003125\,\sigma^{-3}$ to $0.2\, \sigma^{-3}$.   
We compared the simulation results with the empirical Kohlrausch law \cite{Wright}, which states that the molar conductivity, or in our case $K_m/\rho_s$, depends linearly on the square root of the salt concentration $\sqrt{\rho_s}$,
\begin{equation}
  \frac{K_m}{\rho_s} = \frac{ 2e^2 D_I}{k_B T}= A - B \sqrt{\rho_s}, 
\end{equation}
where $A$ and $B$ are fitting parameters. 
Fig. \ref{fig:Km} shows the simulation results and a fit to Kohlrausch law. 
Note that the parameter $A$ is equal to the limiting molar conductivity $[K_m/\rho_s]_{\rho_s \rightarrow 0}$.

In Table \ref{tab:salt}, we list the Debye length, the ratio $R/l_D$ between the colloid radius and the Debye length, the microion diffusion constant, and the medium conductivity for four different salt concentrations. 
The simulations below are performed at these four salt concentrations.

\begin{table}[htbp]
\centering
\begin{spacing}{1.2}
\begin{tabular}{p{1.4cm}p{0.9cm}p{0.7cm}p{1.8cm}p{2.5cm}}
  \hline
  $\rho_s$ [$\sigma^{-3}$] & $l_D$\,[$\sigma$] & $\kappa R$ & $D_I$ [$\sigma^2/\tau$] & $K_m$ [$e^2/(\sigma\varepsilon\tau)]$ \\
  \hline
  0.003125 & 3.57 & 0.84 & 0.68 $\pm$ 0.03 & 0.0042 $\pm$ 0.0002 \\
  0.0125   & 1.78 & 1.68 & 0.63 $\pm$ 0.04 & 0.016  $\pm$ 0.001  \\
  0.05     & 0.89 & 3.36 & 0.50 $\pm$ 0.03 & 0.050  $\pm$ 0.003  \\
  0.2      & 0.45 & 6.73 & 0.27 $\pm$ 0.01 & 0.108  $\pm$ 0.004  \\
  \hline
\end{tabular}
\end{spacing}
\caption{The Debye length $l_D$, the ratio between the colloid radius and the Debye length $\kappa R$, the microion diffusion constant $D_I$, and the medium conductivity $K_m$ for different salt concentrations.}
\label{tab:salt}
\end{table}

To map the simulation units to real physical numbers, we use an aqueous solution of KCl as a reference system. 
The energy unit $\varepsilon$ is $k_BT=4.1\times 10^{-21}\,\mathrm{J}$ at room temperature $T=298\,\mathrm{K}$.
We set the Bjerrum length $l_B=1.0\,\sigma$, thus the length unit $\sigma$ corresponds to the Bjerrum length of water at room temperature, $l_B=0.71\,\mathrm{nm}$.
A salt concentration $0.2\,\sigma^{-3}$ in simulation translates to an experimental value of $912\,\mathrm{mol/L}$.
We further equate the diffusion constant of microions at zero concentration ($D^0_I=0.75 \,\sigma^2/\tau$) to that of an electrolyte solution \cite{Padding2006}.
The diffusion constant for K$^{+}$ and Cl$^-$ differ slightly ($D_{\rm K^+}=1.96\times 10^{-9}\,\mathrm{m^2s^{-1}}$ and $D_{\rm Cl^-}=2.03\times 10^{-9} \,\mathrm{m^2s^{-1}}$) \cite{Hill2003,Ohshima}, so we use the value $2.0\times 10^{-9}\, \mathrm{m^2s^{-1}}$ to obtain that one simulation time unit corresponds to real time $1.91\times 10^{-10}\,\mathrm{s}$. 
Thus the frequency $f=0.1\,\tau^{-1}$ corresponds to an experimental frequency 523 MHz.

\section{Results and Discussion}
\label{sec:results}

In this section, we report simulations for a single uncharged colloid of radius $R=3.0\,\sigma$ in solutions with different salt concentrations. 
We systematically investigate the effect of varying the frequency of the external AC field, from $f=10^{-3}$ to $2.0\,\tau^{-1}$. 
A separate simulation with a constant electric field is also performed to provide the reference at the low-frequency limit. 
The amplitude of the AC field is chosen in the linear region, $E_0 = 0.5\, \varepsilon/(\sigma e)$. 
The dipole moment is calculated by counting the microions around the colloid,
\begin{equation}
  \mathbf{p}=\sum q_i (\mathbf{r}_i - \mathbf{r}_c),
\end{equation}
where $\mathbf{r}_i$ is the position of microion with charge $q_i$.
Since the total charge is zero, the dipole moment does not depend on the reference point and we use the colloid center $\mathbf{r}_c$. 
We also need to choose a cutoff for the measurement and in this work we count the contribution from microions which have a distance less than $3\,l_D$ from the colloid surface.
We have verified that the results do not change significantly when a larger cutoff is used.
The time series of the dipole moment is collected during the simulation over $2 \times 10^6$ time steps. 
The complex polarizability $\alpha(\omega)$ is obtained by applying a Fourier transform to the time series of the dipole moment. 

\begin{figure*}[htbp]
  \centering
  \includegraphics[width=0.9\columnwidth]{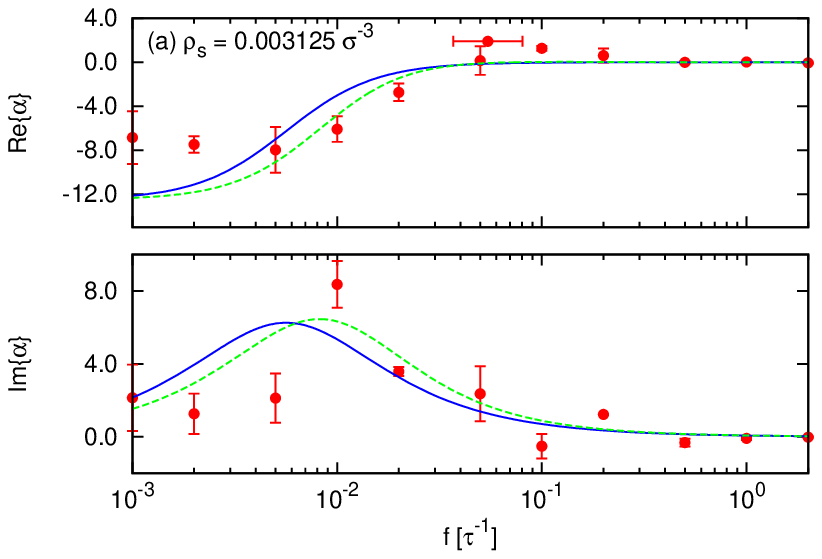}
  \includegraphics[width=0.9\columnwidth]{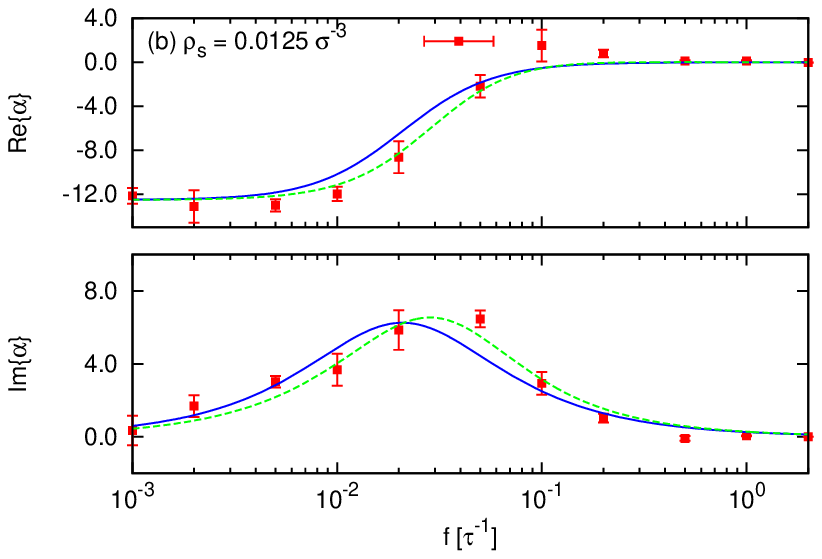}
  \includegraphics[width=0.9\columnwidth]{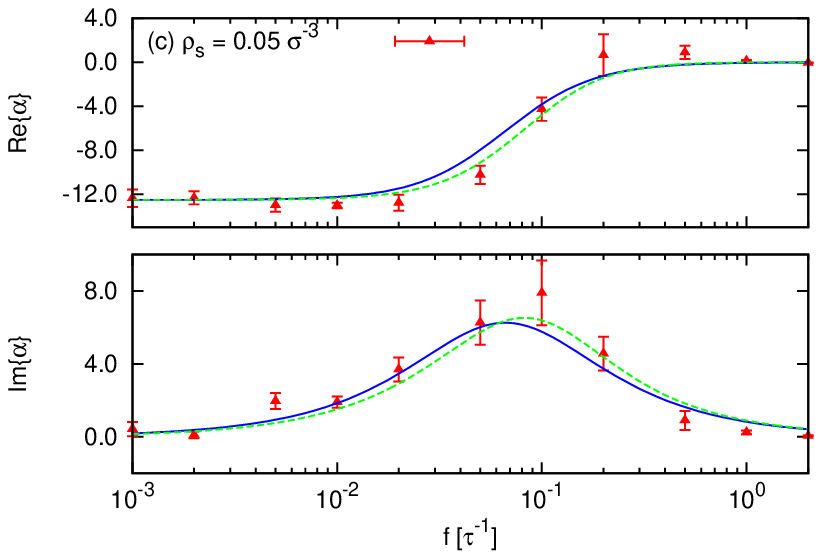}
  \includegraphics[width=0.9\columnwidth]{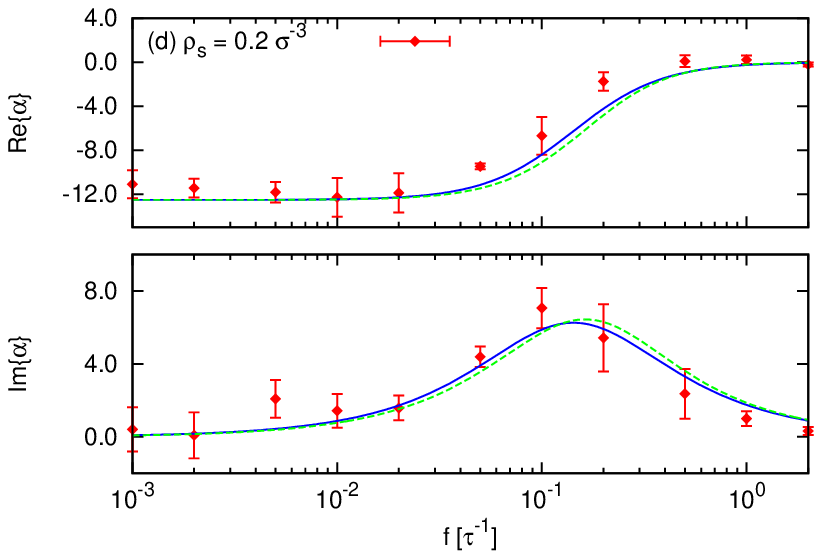}
  \caption{Real and imaginary components of the complex polarizability $\alpha(\omega)$ of an uncharged particle as a function of the frequency of AC field. The field strength is set in the linear region $E_0=0.5\,\varepsilon/(\sigma e)$. The salt concentrations are (a) $\rho_s=0.003125\, \sigma^{-3}$, (b) $\rho_s=0.0125\, \sigma^{-3}$, (c) $\rho_s=0.05\, \sigma^{-3}$, (d) $\rho_s=0.2\, \sigma^{-3}$. The symbols are simulation results. The solid lines give the prediction from the Maxwell-Wagner theory with an effective radius $R_{\rm eff}=2.92\,\sigma$. The dashed lines show the results of Dhont and Kang \cite{Dhont2010}.}
  \label{fig:alpha}
\end{figure*}

Fig. \ref{fig:alpha} shows the simulation results for four different salt concentrations (see Table \ref{tab:salt}), and corresponding Debye lengths ranging from $0.84\,\sigma$ to $6.73\,\sigma$. 
Thus the Debye lengths are comparable to the colloidal radius ($R=3.0\,\sigma$). 
The real part of the polarizability  ${\rm Re}\{\alpha\}$ shows the in-phase component of the dipole moment with respect to the external AC field, while the imaginary part ${\rm Im}\{\alpha\}$ gives the out-of-phase contribution. 
The polarizability in the low-frequency limit has a negative real part, indicating that the induced dipole is anti-parallel to the applied field, in agreement with the theoretical prediction. 
In the opposite limit of high frequency, the colloid and microions can no longer respond to the field, thus both ${\rm Re}\{\alpha\}$ and ${\rm Im}\{\alpha\}$ converge to zero. 

The solid lines in Fig. \ref{fig:alpha} are predictions from Maxwell-Wagner theory. The limiting behavior of the polarizability at frequency $f \rightarrow 0$ was obtained by an independent run in a DC field and used to calibrate the effective colloidal radius. 
Except for the case of the lowest salt concentration $\rho_s=0.003125\,\sigma^{-3}$ (Fig. \ref{fig:alpha}(a)), the low-frequency behavior for different salt concentration gives an effective radius $R_{\rm eff}=2.92 \pm 0.07\, \sigma$, which is reasonably close to the ``physical'' radius $R=3.0\,\sigma$.
Having set the effective colloidal radius, we can calculate the Maxwell-Wagner prediction without further fit parameters. 
The resulting curves are in good agreement with the simulation data. 
Most notably, the theory predicts a crossover between two regions which is recovered in the simulation at roughly the predicted crossover frequency. 
The main discrepancy occurs at low salt concentration $\rho_s=0.003125 \,\sigma^{-3}$ (Fig. \ref{fig:alpha}(a)), where the low-frequency limit of the ${\rm Re}\{ \alpha \}$ is higher than what the theory predicts (with the assumption $R_{\rm eff}=2.92\, \sigma$).  
This is probably due to the finite size of the simulation box, as the Debye length for low salt concentration is quite large, and the resulting cutoff for calculating the dipole moment is about half the size of the simulation box. 
In this case, the interaction between periodic images of the colloid may become important. 
We have qualitatively investigated the effect of finite size at higher salt concentration (smaller $l_D$) by carrying out test runs of smaller systems.
Finite size effects were found to affect the complex polarizability quantitatively, \emph{i.e.}, the value of $\mathrm{Re}(\alpha)$ at low frequency and the peak value of $\mathrm{Im}(\alpha)$.
For linear system sizes $L > 6 l_D$, however, the transition frequency stays about the same.

In the high frequency regime, the inertia effect of microions and solvent also plays a role.
When microions with finite mass are immersed in a viscous fluid, their response to the external field depends on the frequency. 
The characteristic time scale is $t_I=m_I/\lambda_I$, where $\lambda_I=k_BT/D_I$ is the friction constant of microions. 
For solvents, there are two relevant time scales.
One is the kinetic time $t_{\nu}=\sigma/\nu$, defined as the time for momentum to diffuse over the characteristic length $\sigma$, and $\nu$ is the kinetic viscosity of the solvent. 
Because the DPD solvent is compressible, the sound propagation will have an effect on the characteristic time scale of the sonic time $t_{cs}=\sigma/c_s$, where $c_s$ is the speed of sound. 
We do not have a measurement value for the speed of sound, but from the velocity autocorrelation function (inset of Fig. \ref{fig:r2}) we estimate the time of acoustic momentum transport to be  of order unity.
In our simulation, $t_I \sim 3.7 \, \tau$ for $\rho_s=0.2\sigma^{-3}$, and $t_{\nu} \sim 2.4 \,\tau$. 
All these time scales are of order unity, therefore the system will be affected by inertia effects in the high frequency regime $f>1\, \tau^{-1}$. 
At low salt concentrations, the transition frequency is well below this value.
At higher salt concentration, the transition frequency increases and inertia effects  come into play, leading to a slight reduction of the transition frequency.

The dashed curves in Fig. \ref{fig:alpha} show the prediction of Dhont and Kang \cite{Dhont2010} (also see Eq. (\ref{eq:alpha_dhont})). 
The difference between the two theoretic predictions is most prominent at low salt concentration, where the Dhont-Kang theory predicts a higher transition frequency than the Maxwell-Wagner theory.
The polarization charges are distributed over a layer of width $\sim l_D$ (Debye length), and for small salt concentration, the Debye length becomes large. 
Thus the spatial variation of the polarization charges plays a more significant role when $l_D \gg R$.   
At the other limit of  high salt concentration, the effect becomes less important, and Eq. (\ref{eq:alpha_dhont}) can be shown to converge to the 
Maxwell-Wagner result, Eq. (\ref{eq:alpha_mw}).
This can also be seen in Fig. \ref{fig:alpha}(d), where the two curves predicted by the different theories differ only slightly. 
It should be noted that the difference between the curves are smaller than the error of the simulation data at all salt concentrations. 
Thus the data on the polarizability cannot be used to assess the validity of the Dhont-Kang theory. 

\begin{figure}[htbp]
  \centering
  \includegraphics[width=0.9\columnwidth]{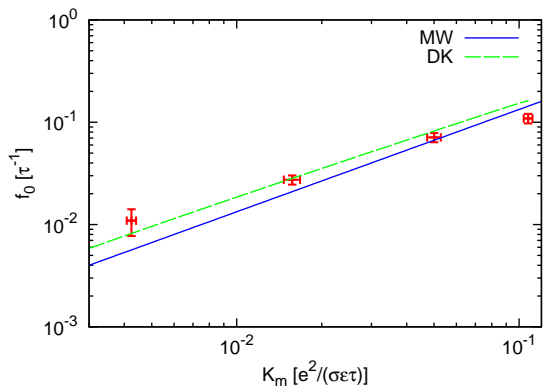}
  \caption{The transition frequency $f_0$ as a function of the medium conductivity $K_m$. The transition frequency is calculated by fitting the imaginary part of the polarizability ${\rm Im}\{\alpha\}$ by a Lorentzian. The solid and dashed lines are predictions from Maxwell-Wagner theory and Dhont-Kang results, respectively.}
  \label{fig:f0}
\end{figure}

\begin{figure}[htbp]
  \centering
  \includegraphics[width=0.9\columnwidth]{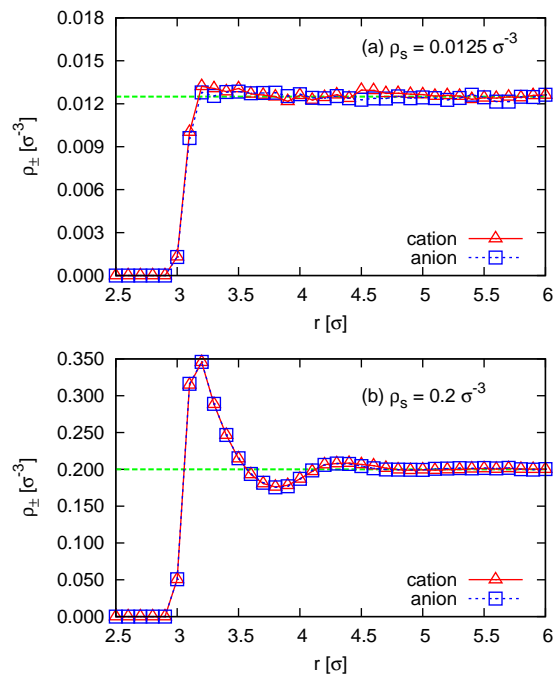}
  \caption{The cation and anion densities as a function of the distance to the colloid center. Densities for two different salt concentrations are shown here: (a) $\rho_s=0.0125\, \sigma^{-3}$ and (b) $\rho_s=0.2\, \sigma^{-3}$. The dashed lines show the bulk value.}
  \label{fig:layering}
\end{figure}

Next we show the transition frequency $f_0$ of the dielectric response as a function of the medium conductivity $K_m$ in Fig. \ref{fig:f0}.
The transition frequency is calculated by fitting the imaginary part of the polarizability ${\rm Im}\{\alpha\}$ by a Lorentzian. 
Also shown are the prediction from the Maxwell-Wagner theory (cf. Eq. (\ref{eq:w_mw})) and numerical results based on the Dhont-Kang theory. 
When the effect of polarization charge distribution is considered, the transition occurs at higher frequency than predicted by the Maxwell-Wagner theory.
At high salt concentration (or large medium conductivity), the predictions of the two theories come close, but the transition frequency obtained from the simulations is lower than the prediction from both theories. 
The inertia effect mentioned above may contribute to this discrepancy due to the fact that $f_0$ approaches $1\,\tau^{-1}$ for high salt concentrations.
Furthermore, an analysis of the salt density distribution indicates that density oscillations appear near the surface for high salt concentration. 
In Fig. \ref{fig:layering}, the averaged ion densities are shown for two different salt concentrations. 
For low salt concentration ($\rho_s=0.0125\, \sigma^{-3}$, Fig. \ref{fig:layering}(a)), the ion density profile is monotonous and rises abruptly from near zero to the bulk value at the colloid surface ($r \le 3\,\sigma$). 
For high salt concentration ($\rho_s=0.2\, \sigma^{-3}$, Fig. \ref{fig:layering}(b)), the density profile exhibits oscillations close to the surface. 
This layering effect of the microions near the colloidal surface at high salt densities is due to excluded volume effects between ions and may contribute to the reduced transition frequency in simulations. 

At low salt concentration (or small medium conductivity), the Debye length is large compared to the colloidal radius, thus the effect of polarization charge is significant. 
The simulation results agree better with the Dhont-Kang result in this regime. 
The two theories differ in their assumptions how the polarization charges are distributed. 
In the Maxwell-Wagner theory, the induced charges only appear at the interface, {\it i.e.} the colloidal surface, and a Laplace equation is assumed to hold in the solution region. 
In the electrokinetic theory, a Poisson equation is applied in the solution region, and the Dhont-Kang results predict a distribution of polarization charges with the characteristic length being the Debye length. 
Simulations can provide useful information about the charge density around the interface and verify the predictions from different theories.
We shall demonstrate the existence of spatially varying polarization charges by two methods.

\begin{figure*}[htbp]
  \centering
  \includegraphics[width=1.6\columnwidth]{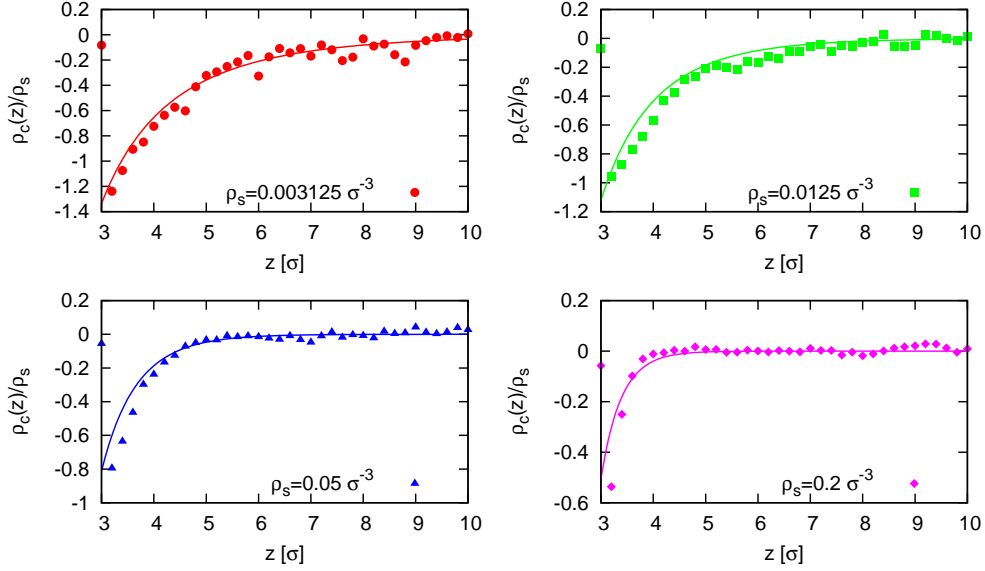}
  \caption{The charge density $\rho_c=e(\rho_+-\rho_-)$ as a function of the distance from the colloid center along the direction of the external field. The charge densities have been normalized using the bulk salt concentration $\rho_s$. The results for four different salt concentrations are shown here (see Table \ref{tab:salt}). The curves are predictions from Eq. (\ref{eq:ion_c}).}
  \label{fig:ion_c}
\end{figure*}

First, we can directly measure the local charge density of microions in the simulation.  
This is more easily implemented for a static electric field, as the charge distribution remains stationary and an average over long time can provide better signal-to-noise ratio.
Fig. \ref{fig:ion_c} shows the averaged charge density $\rho_c(z)$ as a function of the distance $z$ from the center of the sphere along the direction of the external field. 
The curves are calculated using Eq. (\ref{eq:ion_c}). 
The simulation demonstrates directly the spatial distribution of the polarization charges.
There is an excess of negatively charged ions on the front end of the colloid with respect to the external field, and the charge density decays to the bulk value over a distance characterized by the Debye length. 
For low salt concentration, the extent of the distribution is quite large, while for high salt concentration, it decays quickly to zero. 
The prediction from the electrokinetic theory and the simulation result agree quite well with each other.

\begin{figure*}[htbp]
  \includegraphics[width=2.0\columnwidth]{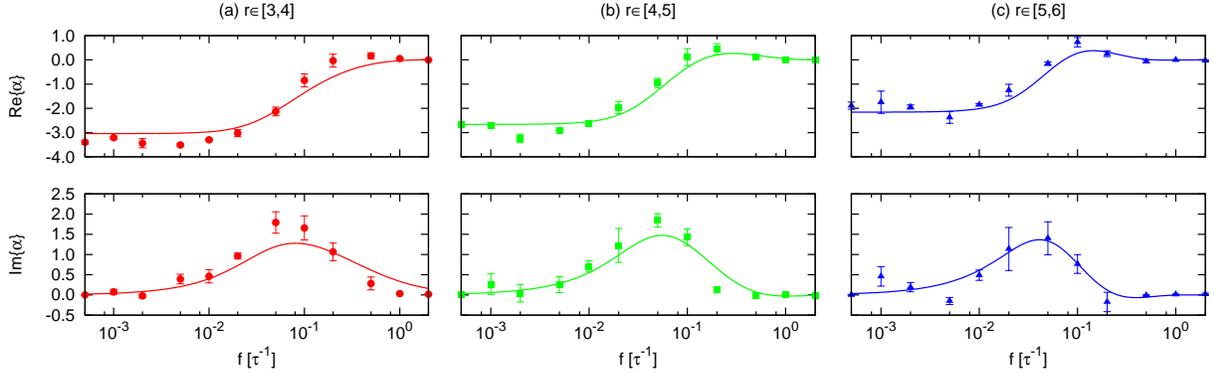}
  \caption{Contribution to the polarizability from microions located in a spherical shell at distances (a) $r \in [3, 4]$, (b) $[4, 5]$, and (c) $[5, 6]$, respectively. The salt concentration is $\rho_s=0.0125\, \sigma^{-3}$ and the corresponding Debye length is $l_D=1.78\,\sigma$. The curves are predictions from Eq. (\ref{eq:p_shell}).}
  \label{fig:dif}
\end{figure*}

Secondly, we can divide the space surrounding the colloid into spherical shells centered at the colloid and compute the contribution to the polarization from each shell.
If the polarization charges are spatially distributed instead of being localized at the interface, the microions in the shell far away from the interface will contribute to the total polarizability.  
Using Eq. (\ref{eq:p_shell}), we can compute this contribution from the electrokinetic theory.
The results for $\rho_s=0.0125\, \sigma^{-3}$ are shown in Fig. \ref{fig:dif} for three different shells. 
The simulation results are in good agreement with the prediction, even the slight overshoot near the crossover region for the shell with large radius is reproduced. 
The contribution diminishes when we move away from the colloidal surface and reduces to noise level when the shell radius is more than three times of the Debye length.

\section{Summary}
\label{sec:summary}

We have carried out mesoscopic molecular dynamics simulations of an uncharged colloidal particle in electrolyte solution under alternating electric fields, for different salt concentrations. 
We have taken full account of the hydrodynamic interaction with thermal fluctuations, using Dissipative Particle Dynamics, and the electrostatic interactions, using a Particle-Particle-Particle Mesh method.
 
We obtained information about the dielectric response of the particle to a weak external AC field by computing the complex polarizability of the colloid.
We systematically investigated the effect of the AC field frequency and the bulk salt concentration.
The simulation results were compared with the predictions of the Maxwell-Wagner theory \cite{Maxwell1954,Wagner1914} and from electrokinetic theory \cite{Dhont2010,Bonincontro1980,Garcia1985}. 
For the total polarizability, both theories show good agreement with the simulation, but the electrokinetic theory provides more information about the distribution of the polarization charges. 

There are some subtle differences between the electrokinetic model used in ref. \cite{Dhont2010} and our simulation model. 
In ref. \cite{Dhont2010}, the convective contribution in the Nernst-Planck equation is relatively small for uncharged colloids, thus the convective term is omitted. 
This is necessary to make analytical progress; it would be a formidable task to solve all equations including the Navier-Stokes equations. 
In the simulation model, the hydrodynamic interaction is included through the dissipative particle dynamics. 
The good agreement between the simulation and theory indicates that, for uncharged colloids, the hydrodynamics is not important. 
However, this is not true for charged colloids, where the contribution from the electrophoretic motion of the colloid and the electro-osmotic flow alter the results significantly.  

Secondly, the salt microions are considered to be point-like particles in the electrokinetic equations, which are introduced through the ionic concentration. 
This treatment corresponds to a mean-field type approach and neglect the ion-ion correlations.
The effect of ion-ion correlations is small for the system of low salt concentration, but becomes important at high salt concentration.
In the simulation, we used a particle-based approach, and microions are modeled as spherical beads with finite size. 
As a consequence, we observed the layering effect of the microions near the colloid surface, which cannot be described by a theory that neglect ion-ion correlations.
It will be interesting to compare the simulation results to the electrokinetic model which includes the ion size effect \cite{Roa2011a,Roa2011b}.

\begin{acknowledgments}
This work was funded by the Deutsche Forschungsgemeinschaft (DFG) through the SFB-TR6 program ``Physics of Colloidal Dispersions in External Fields''. 
Computational resources at John von Neumann Institute for Computing (NIC J\"ulich), High Performance Computing Center Stuttgart (HLRS) and JGU Mainz (MOGON) are gratefully acknowledged.
\end{acknowledgments}

\appendix

\section{Maxwell-Wagner theory}
\label{app:mw}

In this appendix, we give a short introduction to Maxwell-Wagner theory. More detailed information can be found in refs. \cite{RSS,Maxwell1954,Wagner1914}.

When loss is present, the dipole moment of a particle immersed in fluids
exhibits a phase lag with respect to the external AC field $\mathbf{E_0} \exp(i\omega t)$. 
A complex dipole moment can be written as 
\begin{equation}
  \label{eq:dipole}
  \mathbf{p} = 4 \pi \epsilon_m K(\omega) R^3 \: \mathbf{E}_0 ,
\end{equation} 
where the Clausius-Mossotti factor $K(\omega)$ is a complex number containing both the magnitude and the phase information about the effective dipole moment.  
In the Maxwell-Wagner theory, it has the form
\begin{equation}
  \label{eq:cm2}
  K (\epsilon^*_p, \epsilon^*_m) 
    = \frac{ \epsilon^*_p - \epsilon^*_m }{ \epsilon^*_p + 2\epsilon^*_m }, 
\end{equation}
where $\epsilon^*_p$ and $\epsilon^*_m$ are the complex dielectric constants of
the particle and the medium, respectively.  
They are defined as
\begin{equation}
  \epsilon^*_p = \epsilon_p + \frac{ K_p }{i\omega}, \quad
  \epsilon^*_m = \epsilon_m + \frac{ K_m }{i\omega},
\end{equation}
where $\epsilon$ (without the star) and $K$ are the permittivity and conductivity, respectively. 
The real and imaginary parts of the complex Clausius-Mossotti factor can be expressed as
\begin{eqnarray}
  {\rm Re}\{ K^* \} &=& \frac{ K_p - K_m }{K_p + 2K_m} \frac{1 + \omega \tau_0
    \tau_{\rm mw} }{ 1+ \omega^2 \tau_{\rm mw}^2}, \\
  {\rm Im}\{ K^* \} &=& \frac{ K_p - K_m }{K_p + 2K_m} \frac{\omega (\tau_0 -
    \tau_{\rm mw}) }{ 1+ \omega^2 \tau_{\rm mw}^2}, \\
  \tau_0 &=& \frac{ \epsilon_p - \epsilon_m }{K_p-K_m}, \\
  \tau_{\rm mw} &=& \frac{ \epsilon_p + 2 \epsilon_m} {K_p + 2K_m}.
\end{eqnarray}
For $\epsilon_m=\epsilon_p=\epsilon$ and $K_p=0$, the Clausius-Mossotti factor is reduced to
\begin{eqnarray}
  \label{eq:Re_K2}
  {\rm Re}\{ K^* \} &=& - \frac{1}{2} \: \frac{1}{ 1+ (\omega / \omega_{\rm mw})^2}, \\ 
  \label{eq:Im_K2}
  {\rm Im}\{ K^* \} &=& \frac{1}{2} \: \frac{\omega / \omega_{\rm mw}}{ 1+ ( \omega / \omega_{\rm mw})^2},
\end{eqnarray}
with $\omega_{\rm mw} = 2 K_m/(3 \epsilon)$.
The induced dipole moment is 
\begin{equation}
  \label{eq:p_mw}
  \mathbf{p} = -2 \pi \epsilon \mathbf{E}_0 R^3 \frac{ 1 - i(\omega/\omega_{\rm mw}) }{ 1 + (\omega/\omega_{\rm mw})^2}. 
\end{equation}
This corresponds to the polarizability given in Eq. (\ref{eq:alpha_mw}).

\section{Dhont-Kang theory}
\label{app:dhont}

Dhont and Kang \cite{Dhont2010} derived a formula for the complex response function of the polarization charge density.
Here for the purpose of comparison with simulations, we derive two equations (\ref{eq:p_shell}) and (\ref{eq:ion_c}) for our system from their results. 
The system consists of an uncharged non-conducting colloid immersed in an electrolyte solution with the same permittivity as the colloid core. 
The external electric field has a form of $\mathbf{E} = \mathbf{E}_0 \exp( i \omega t)$ and the charge density can then be written as
\begin{equation}
	\rho_c (\mathbf{r}, t) = R(\mathbf{r}, \omega) \exp( i \omega t),
\end{equation}	
where $R$ is the complex-valued response function for the polarization charge density. 
The response function takes the form
\begin{equation}
	\label{eq:R}
	R(\mathbf{r}, \omega) = R_0 (\omega) \mathbf{E}_0 \cdot \nabla \frac{ \exp(-sr) }{r}, 
\end{equation}
with 
\begin{eqnarray}
	R_0(\omega) &=& \frac{ \epsilon \kappa^2 R^3 \exp(sR) } { 2 + 2sR + (sR)^2 - \frac{1}{3} (\kappa R)^2 }, \\
	s &=& \kappa ( x + i y ), \label{eq:s_app} \\
	x &=& \frac{1}{\sqrt{2}} \left[ 1 + (1+\Lambda^2)^{1/2} \right]^{1/2}, \\ 
	y &=& \frac{1}{\sqrt{2}} \left[ -1 + (1+\Lambda^2)^{1/2} \right]^{1/2}, \\ 
	\Lambda &=& \frac{\omega}{ \kappa^2 D_I} \label{eq:l_app} .
\end{eqnarray}

From the response function $R$, the contribution to the dipole moment from the shell $r_1 < r < r_2$ can be readily calculated
\begin{eqnarray}
	&& \mathbf{p}(\omega | r_1,r_2) = \int_{r_1}^{r_2} \mathrm{d} \mathbf{r} \, \mathbf{r} R(\mathbf{r}, \omega)  \nonumber \\
        \label{eq:p_shell}
	&& = - 4\pi \mathbf{E}_0 R_0 (\omega) \frac{ \exp(-sr) }{s^2} \left[1 + sr + \frac{1}{3}(sr)^2 \right] \Big|_{r_1}^{r_2}  \quad \quad
\end{eqnarray}
The total dipole moment is then obtained by taking $r_1 \rightarrow R$ and $r_2 \rightarrow \infty$,
\begin{eqnarray}
	&& \mathbf{p}(\omega) = \mathbf{p}(\omega | R, \infty) \nonumber \\
	&& = -4\pi\epsilon \mathbf{E}_0 R^3 \frac{\kappa^2}{s^2} \frac{ 1 + sR + \frac{1}{3} (sR)^2 }{ 2 + 2sR + (sR)^2 - \frac{1}{3} (\kappa R)^2 } \quad \quad
        \label{eq:dk_dipole}
\end{eqnarray}
This is Eq. (30) in ref. \cite{Dhont2010}, and it can be shown to be in agreement with Eq. (19) in ref. \cite{Garcia1985}.

Eq. (\ref{eq:dk_dipole}) can be rewritten using (\ref{eq:s_app})-(\ref{eq:l_app}) and taking the limit $sR \rightarrow \infty$,
\begin{equation}
  \mathbf{p}(\omega) = -2\pi\epsilon \mathbf{E}_0 R^3 \frac{1}{1 + \frac{3}{2} \Lambda i}.  
\end{equation}
The factor $\frac{3}{2}\Lambda$ in the denominator can be expressed in terms of $K_m$ using Eq. (\ref{eq:l_app}), (\ref{eq:Km}) and (\ref{eq:Debye}),
\begin{equation}
  \frac{3}{2}\Lambda = \frac{3 \omega}{2 \kappa^2 D_I} = \frac{3\omega \epsilon}{2 K_m} = \frac{\omega}{\omega_{\rm mw}}.
\end{equation}
This is exactly the result from the Maxwell-Wagner theory (cf. Eq. (\ref{eq:p_mw})). 

For a constant electric field, the imaginary part of the response function vanishes. 
The real part of $R$, which in this case ($\omega=0$) is just the charge density $\rho_c$,  as a function of the distance from the center of the colloid along the direction of the external field is
\begin{eqnarray}
  &&\rho_c(z) = {\rm Re} \{ R (z, \omega=0) \} \nonumber \\
  \label{eq:ion_c}
  &&= - \epsilon E_0 \frac{ \kappa^2 R^3 }{ 2 + 2\kappa R + \frac{2}{3} (\kappa R)^2 } \frac {1+\kappa r}{r^2} e^{ -\kappa (r-R) }. \quad \quad \quad 
\end{eqnarray}

\bibliography{uncharged}

\end{document}